\documentclass[conference]{IEEEtran}
\IEEEoverridecommandlockouts
\usepackage{cite}
\usepackage{amsmath,amssymb,amsfonts}
\usepackage{algorithm}
\usepackage{algorithmic}
\usepackage{graphicx}
\usepackage{multirow}
\usepackage{subfigure}
\usepackage{textcomp}
\usepackage{xcolor}
\def\BibTeX{{\rm B\kern-.05em{\sc i\kern-.025em b}\kern-.08em
    T\kern-.1667em\lower.7ex\hbox{E}\kern-.125emX}}
\begin{document}
\title{Loss Prediction: End-to-End Active Learning Approach For Speech Recognition}

\author{
\IEEEauthorblockN{Jian Luo, Jianzong Wang*\thanks{*Corresponding author: Jianzong Wang, jzwang@188.com}, Ning Cheng, Jing Xiao}
\IEEEauthorblockA{\textit{Ping An Technology (Shenzhen) Co., Ltd.}}
\textit{Shenzhen, China}}

\maketitle

\begin{abstract}
End-to-end speech recognition systems usually require huge amounts of labeling resource, while annotating the speech data is complicated and expensive. Active learning is the solution by selecting the most valuable samples for annotation. In this paper, we proposed to use a predicted loss that estimates the uncertainty of the sample. The CTC (Connectionist Temporal Classification) and attention loss are informative for speech recognition since they are computed based on all decoding paths and alignments. We defined an end-to-end active learning pipeline, training an ASR/LP (Automatic Speech Recognition/Loss Prediction) joint model. The proposed approach was validated on an English and a Chinese speech recognition task. The experiments show that our approach achieves competitive results, outperforming random selection, least confidence, and estimated loss method.
\end{abstract}


\section{Introduction}
\label{sec:intro}

Over the past decades, end-to-end neural networks have achieved remarkable results in automatic speech recognition tasks. Among these models, CTC~\cite{graves2006connectionist}, encoder-decoder~\cite{dong2018speech} and CTC-attention hybrid~\cite{watanabe2017hybrid, kim2017joint} architectures attract amounts of attention. The key success of the CTC-attention hybrid model is that they use a shared-encoder representation trained by both CTC and attention loss simultaneously. The CTC object limits the alignment of the encoder, thus helping the attention object to obtain performance improvement and fast convergence. However, these end-to-end models often suffer from low computational efficiency because of complicated network structures. They also require a large
amount of transcribed speech data for training. It is known that labeling speech data is costly and time-consuming. Labeling of audio can take about $10$ times longer than the actual audio duration~\cite{semi2005zhu}. Therefore, optimizing training efficiency is necessary for improving the performance of end-to-end ASR systems.

Active learning (AL) is the method describing how to utilize the labeling resources efficiently. AL improves the training efficiency by selecting the most valuable data samples from an unlabeled database for annotation.
The selected samples for training are mostly unfamiliar and uncertain to the current model. The most common active learning methods can be categorized into three categories: (1) pool-based sampling, (2) stream-based sampling, and (3) query synthesis. The sampling methods use decision strategies to select the most informative samples, while query-synthesis methods apply generative models to generate training instances based on unlabeled pools~\cite{miller2014adversarial,mahapatra2018efficient}. The sampling strategies have been successfully applied in many natural language processing tasks, such as information extraction, named entity recognition, text categorization, etc~\cite{literature2009olsson}. The synthesis methods use the concept that it is feasible to generate highly informative artificial training instances. However, these synthesis methods have not been proved stable on textual or speech data~\cite{schumann2019active}. Therefore, we mainly focus on the sampling strategies in this work.

Among the sampling strategies, (1) uncertainty-based~\cite{david1994heterogeneous,yu2010active,tur2005combining,gal2015bayesian}, (2) diversity-based~\cite{sener2017active,nguyen2004active}, and (3) expected-model-change-based~\cite{huang2016activelf,yuan2019gradient} algorithm are three major approaches. Uncertainty is defined as the classifiers to select the least uncertain data points to label~\cite{cohn1994improving}. Posterior probability and entropy of predicated classes are common standards for measuring uncertainty in~\cite{lewis1994sequential,joshi2009multi,wang2016cost}. Besides, probability and entropy are also considered as the representations of the Bayesian framework and non-Bayesian framework in uncertainty sampling, respectively~\cite{sinha2019variational,tran2019bayesian}. This sampling method is proved to achieve better performance than random selection in various image-related and textual tasks~\cite{settles2009active,yoo2019learning,hakkani2002active,settles2008analysis,malhotra2019active}. Diversity sampling is also known as the representation-based method. \cite{sener2017active} defines diversity as a Core-Set selection, choosing samples such that a model learned over the selected subset is competitive for the remaining data. They select the unlabeled data based on the distribution of its intermediate feature space~\cite{guo2010active,malhotra2019active}. Furthermore, integrating uncertainty and diversity in selection strategies are also proposed in active learning as presented in~\cite{liu2020integrating}. Apart from previous statistical methods, the expected-model-change-based approach samples the data that causes the largest change to the model in the current training iteration. The expected change can be computed by estimating the expected gradient or expected errors~\cite{settles2008multiple,roy2001toward}. However, this approach is infeasible to be applied in deep neural networks as it is overwhelmingly compute-intensive~\cite{yoo2019learning}.

In summary, our major contributions are the followings:
\begin{itemize}
	\setlength{\itemsep}{0pt}
	\setlength{\parsep}{0pt}
	\setlength{\parskip}{0pt}
	\item Propose to use a loss prediction module on active learning of ASR models, leveraging the loss information computed on all path probabilities.
	\item Design a length-normalized ranking metric of data selection, converting CTC and attention loss into the same scale.
	\item Define an ASR/LP (Automatic Speech Recognition/Loss Prediction) joint model, to enable an end-to-end active learning pipeline.
\end{itemize}

The rest of the paper is organized as follows. Sec.~\ref{sec:related} highlights
the related prior works. Sec.~\ref{sec:method} describes the proposed model architecture of active learning and the algorithm of data selection. Sec.~\ref{sec:exp} reports the experimental results compared with other AL methods. We also discuss the choices on error function and the correlations about ranking metric. In Sec.~\ref{sec:conclu}, we conclude with the summary of the work.

\section{Related Works}
\label{sec:related}

In this paper, we explore the application of uncertainty-based active learning for end-to-end ASR models. For ASR tasks, predicting uncertainty and diversity of samples is difficult because text transcription is a sequence of labels. In recent studies, the probability-based least confident score was explored. The samples with the lowest decoding path probability are considered the most informative data for the current model~\cite{hakkani2002active,settles2008analysis}. However, the least confidence method only considers the most likely decoding path of the audio samples without other path probabilities.
We thought the probabilities of all decoding paths might contribute to evaluating the samples. Therefore, in this paper, we used ``loss'' of unlabeled samples as the ranking metric to select annotated data. Another popular approach is Expected Gradient Length (EGL)~\cite{yuan2019gradient}. EGL selects the samples expected to have the largest gradient length, and approximates the true gradient length through a neural network. However, this method might be sensitive to outliers since outliers usually have a large gradient. Moreover, the gradient predicted network is trained separately from the ASR network, resulting in a non-end-to-end active learning pipeline.

CTC~\cite{graves2006connectionist} and attention~\cite{chan2015listen} loss are most frequently used in end-to-end ASR models. CTC is hard alignment, marginalizing all possible alignments monotonically, while attention is soft alignment, aligning the inputs and outputs via the attention mechanism. Both CTC and attention loss are computed based on all decoding paths and alignments. However, we cannot obtain these losses unless the ground-truth label sequences are provided. The work called pCTC (predicted CTC)~\cite{huang2016activelf} takes the most likely prediction as the ground-truth sequence and calculate its CTC loss by these prediction as loss estimation. Yet the pCTC method might not work, especially in the early stages of model training. Since the model is still not sufficiently trained, and the prediction sequences are quite distinct from the ground-truth labels, which cannot be used in CTC loss calculation. Based on this motivation, we designed a loss prediction module (LP) to estimate the loss, and integrated it into an existing ASR model. Our works were mainly inspired by~\cite{yoo2019learning}. Loss prediction module returns a loss value that estimates the informativeness of unlabeled data. The authors of \cite{yoo2019learning} also demonstrated consistent high achievements in image classification and regression tasks. A similar idea was also proposed by~\cite{pimentel2020deep}, which validated it was practical in image anomaly detection. To our knowledge, the loss prediction has not been applied to speech data yet. In this work, we extend the loss prediction module to hybrid CTC-attention speech recognition models and define a new joint network structure.


\section{Proposed Method}
\label{sec:method}
\begin{figure}[ht]
	\begin{center}
		\centerline{\includegraphics[width=0.8\columnwidth]{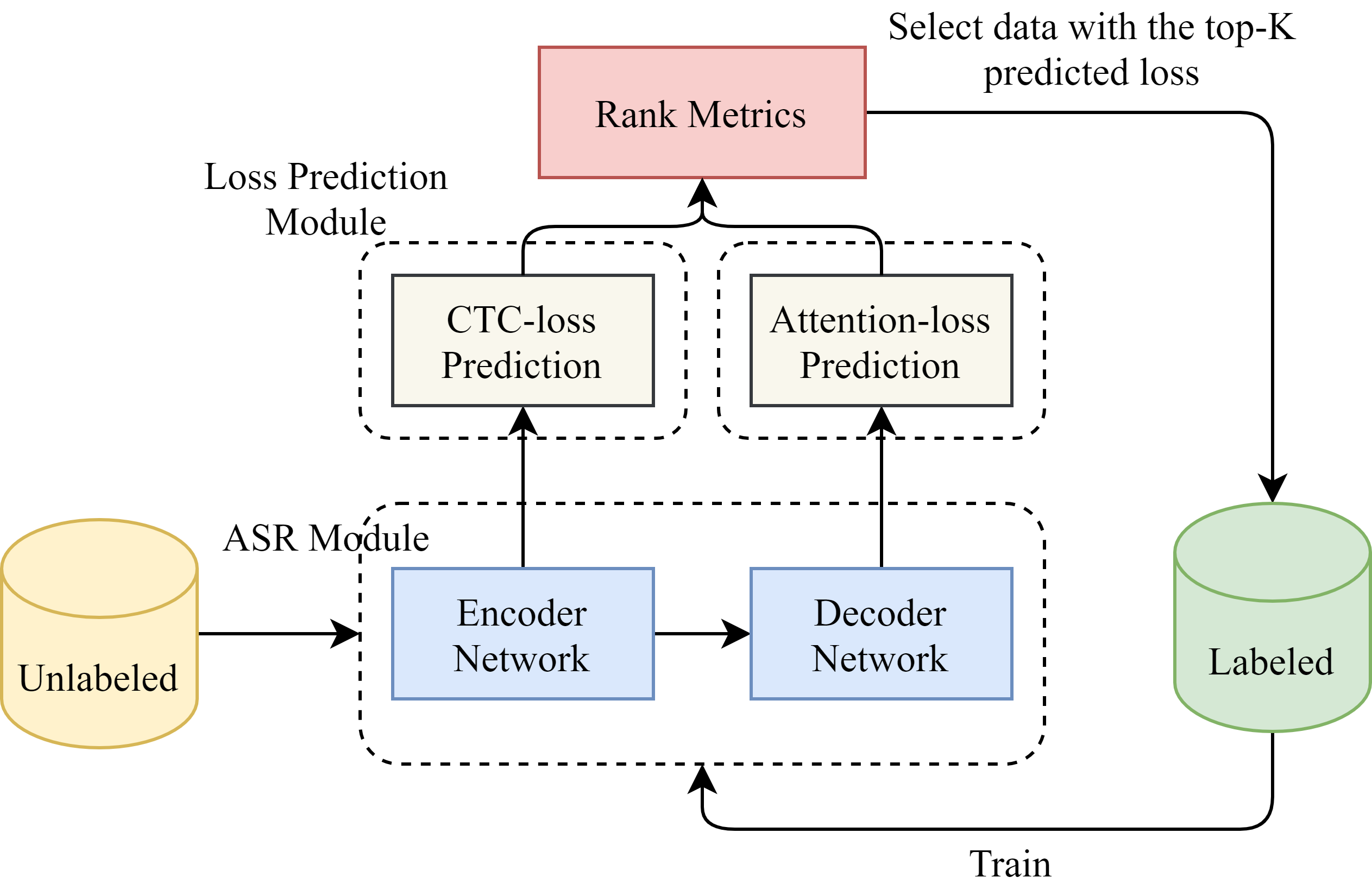}}
		\caption{Active Learning Pipeline with Loss Prediction}
		\label{fig1}
	\end{center}
\end{figure}

In this paper, we propose an active training pipeline for end-to-end speech recognition. As Fig~\ref{fig1} depicts, the overall training pipeline is performed iteratively. It consists of two parts: (1) ASR/LP joint model training, (2) annotated data selection. The ASR/LP joint model contains ASR target task module $\theta_{asr}$ and loss prediction module $\theta_{lp}$. Task module $\theta_{asr}$ conducts the ASR task $\tilde{y}=\theta_{asr}(x)$, where $x$ is the audio sample and $\tilde{y}$ is the predicted text sentence. Loss prediction module predicts the loss $\tilde{\mathcal{L}}=\theta_{lp}(h,g)$, where $h$ and $g$ are the hidden feature maps extracted from task module $\theta_{asr}$.

\subsection{Target Task (ASR) Module}
The hybrid CTC-attention model~\cite{watanabe2017hybrid, kim2017joint} has impressive performance in end-to-end speech recognition tasks. We use the hybrid model as our ASR target task module $\theta_{asr}$. The hybrid CTC-attention model uses CTC objective function as an
auxiliary task to train the attention model encoder with the multi-task learning framework. The hybrid ASR model contains two parts: $\theta_{asr} = [\theta_{encoder},\theta_{decoder}]$. The encoder network $\theta_{encoder}$ is shared with CTC and attention models. The encoder transforms a speech acoustic feature sequence $x = (x_1,x_2,...,x_T)$ to a hidden representation $h = (h_1,h_2,...,h_T)$. The decoder network $\theta_{decoder}$ outputs a text label sequence $\tilde{y} = (\tilde{y}_1,\tilde{y}_2,...,\tilde{y}_S)$ one character at each step.

The CTC loss $\mathcal{L}_{ctc}$ of the ASR module to be minimized is defined as the negative log likelihood of the ground truth text sequence $y = (y_1,y_2,...,y_S)$.
\begin{equation}
h = \theta_{encoder}(x)
\end{equation}
\begin{equation}
\mathcal{L}_{ctc} \triangleq -\ln~P(y|h) = -\ln~\prod_{t}P(y|h_t)
\end{equation}
Where $P(y|h_t)$ is the probability of the corresponding text label at frame step $t$ of encoder output.

The attention model predicts each target label, conditioning on previous labels according to the recursive equation. The attention loss $\mathcal{L}_{attention}$ is computed between predicted sequence $\tilde{y}$ and ground truth sequence $y$ based on the cross-entropy criterion.
\begin{equation}
g_s = \theta_{decoder}(h, \tilde{y}_{1:s-1})
\end{equation}
\begin{equation}
\begin{aligned}
\mathcal{L}_{attention} &\triangleq -\ln~P(y|g) = -\sum_{s}\ln~P(y|g_s) \\
&= -\sum_{s}\ln~P(y_s|h,\tilde{y}_{1:s-1})
\end{aligned}
\end{equation}
Where $g$ represents the output hidden states of decoder network $g = (g_1,g_2,...,g_S)$. $P(y_s|h,\tilde{y}_{1:s-1})$ is the probability of the target label $y_s$ at text sequence step $s$ of decoder output, and $\tilde{y}_{1:s-1}$ is the previous predicted characters.

The target task loss $\mathcal{L}$ of ASR module is calculated through both the CTC loss $\mathcal{L}_{ctc}$ and the attention loss $\mathcal{L}_{attention}$:
\begin{equation}
\mathcal{L} = \lambda~\mathcal{L}_{ctc} + (1-\lambda)~\mathcal{L}_{attention}
\end{equation}
with a scale factor $\lambda: 0 \leq \lambda \leq 1$.

\subsection{Loss Prediction (LP) Module}
The objective of LP module is to learn a good loss prediction model $\theta_{lp}$ for the ranking metric of data selection. Given the training audio sample $x$ and ground-truth text $y$, ASR module can compute target CTC loss $\mathcal{L}_{ctc}$ and attention loss $\mathcal{L}_{attention}$. Therefore, we also design two parts of loss prediction: $\theta_{lp} = [\theta_{ctc},\theta_{attention}]$.

The CTC loss prediction model $\theta_{ctc}$ is defined to predict CTC loss $\tilde{\mathcal{L}}_{ctc}$ by given input audio sample $x$. We can first obtain the encoder hidden features through the ASR encoder $\theta_{encoder}$. Afterwards, $\theta_{ctc}$ takes the last layer hidden states $h$ of $\theta_{encoder}$ as input. Then the predicted CTC loss $\tilde{\mathcal{L}}_{ctc}$ can be computed by:
\begin{equation}
\tilde{\mathcal{L}}_{ctc} = \theta_{ctc}(h)
\end{equation}

Similarly, the attention loss prediction model $\theta_{attention}$ computes the predicted attention loss $\tilde{\mathcal{L}}_{attention}$ by the last layer of decoder hidden states $g$:
\begin{equation}
\tilde{\mathcal{L}}_{attention} = \theta_{attention}(g)
\end{equation}

In order to measure the difference between the true loss and predicted loss, we propose to use error function. The most straightforward error function of predicted loss $\tilde{\mathcal{L}}$ is mean square error (MSE) $\mathcal{E}=(\mathcal{L}-\tilde{\mathcal{L}})^2$. However, we find MSE is not suitable because MSE is the L2 scale, while the target ASR loss $\mathcal{L}$ is the L1 scale. Gradient explosion easily occurs, and the LP module fails in converging when training ASR/LP joint model by adding different scales $\mathcal{L}$ and $\mathcal{E}$ together. To resolve this issue, we use SmoothL1Loss~\cite{girshick2015fast} with threshold factor $\beta$, which is a smooth version between L1 Loss and MSE Loss.

The error function of predicted CTC loss is defined as $\mathcal{E}_{ctc}$, computed by following equation:
\begin{equation}
\mathcal{E}_{ctc} = \begin{cases} 0.5*(\mathcal{L}_{ctc}-\tilde{\mathcal{L}}_{ctc})^2/\beta, & \mbox{if} |\mathcal{L}_{ctc}-\tilde{\mathcal{L}}_{ctc}|<\beta \\ |\mathcal{L}_{ctc}-\tilde{\mathcal{L}}_{ctc}|-0.5*\beta, & \mbox{otherwise}\end{cases}
\end{equation}

Similarly, the error function of predicted attention loss as $\mathcal{E}_{attention}$ is also SmoothL1Loss, computed by $\mathcal{L}_{attention}$ and $\tilde{\mathcal{L}}_{attention}$ with factor $\beta$. We will discuss the details of error function choices in Sec.~\ref{sec:analysis}. Then, the total error of LP module $\mathcal{E}$ is calculated by both $\mathcal{E}_{ctc}$ and $\mathcal{E}_{attention}$, with the scale factor $\lambda$:
\begin{equation}
\mathcal{E} = \lambda~\mathcal{E}_{ctc}+ (1-\lambda)~\mathcal{E}_{attention}
\end{equation}

\subsection{ASR/LP Joint Model Training}

\begin{figure}[ht]
	\begin{center}
		\centerline{\includegraphics[width=\columnwidth]{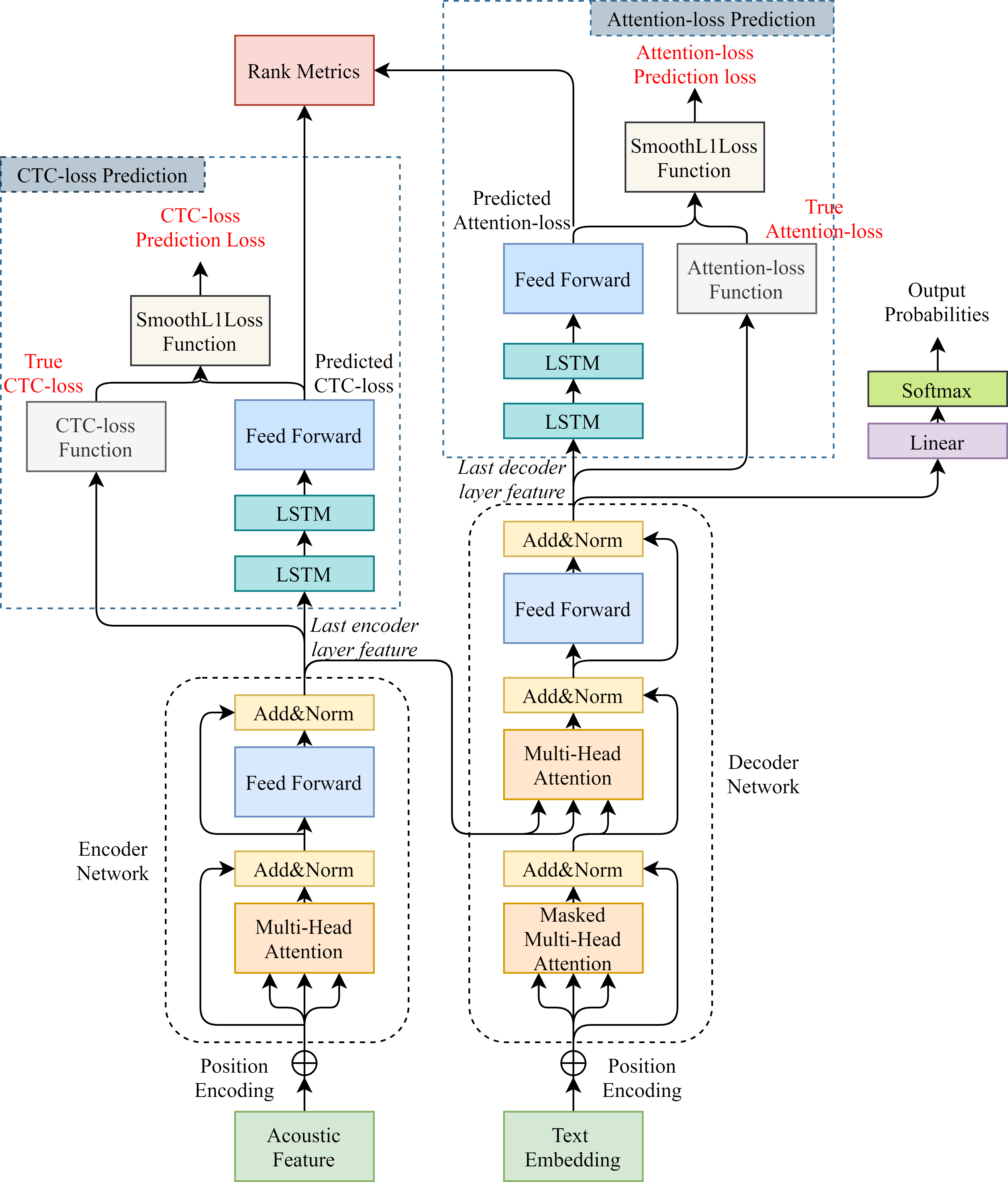}}
		\caption{The Architecture of ASR/LP Joint Model Training}
		\label{fig2}
	\end{center}
\end{figure}

The proposed training method uses a multi-task framework, training ASR module and LP module together. Fig~\ref{fig2} illustrates the overall architecture of joint model training. 

The total cost of joint model training $\mathcal{C}$ is composed of four components: (1) true CTC loss $\mathcal{L}_{ctc}$, (2) true attention loss $\mathcal{L}_{attention}$, (3) CTC loss prediction error $\mathcal{E}_{ctc}$, and (4) attention loss prediction error $\mathcal{E}_{attention}$, with a hyperparameter $\mu$:
\begin{equation}
\begin{aligned}
\mathcal{C} &= \mathcal{L} + \mu~\mathcal{E} \\
&= \lambda~\mathcal{L}_{ctc} + (1-\lambda)~\mathcal{L}_{attention} \\
&+ \mu~\lambda~\mathcal{E}_{ctc} + \mu~(1-\lambda)~\mathcal{E}_{attention}
\end{aligned}
\end{equation}
The ASR and LP module are joint training on the total cost $\mathcal{C}$, using the same labeled dataset at each training iteration.

Compared with the method training ASR model and LP model separately, joint training composes an end-to-end active learning pipeline. Besides, the joint model can activate the ASR module and LP module to conduct the speech recognition and loss prediction receptively, without coupling with each other.

\subsection{Select Annotated Data}
Algorithm~\ref{algorithm1} shows the active learning algorithm for selecting annotated data. The overall training pipeline is an iterative procedure. The algorithm starts with initially unlabeled dataset $D_{u}$ and labeled dataset $D_{l}$. At each iteration of model training, $K$ samples with top-ranking metrics are selected from $D_{u}$. For each sample $x$, the ranking metric $R_{x}$ is computed as:
\begin{equation}
\label{rank_metric}
\begin{aligned}
R_{x} &= \theta_{lp}(x) \\
&= \lambda~\frac{\theta_{ctc}(h)}{T} + (1-\lambda)~\frac{\theta_{attention}(g)}{S} \\
&= \lambda~\frac{\tilde{\mathcal{L}}_{ctc}}{T} + (1-\lambda)~\frac{\tilde{\mathcal{L}}_{attention}}{S}
\end{aligned}
\end{equation}

We normalize the $\tilde{\mathcal{L}}_{ctc}$ by acoustic frames length $T$, and $\tilde{\mathcal{L}}_{attention}$ by text sequence length $S$, to convert them into the same scale. Otherwise, the longer audio samples are always selected, because they have larger loss value. Then the selected dataset $D_{a}$ are annotated by human experts. At next training iteration, the ASR/LP joint model is trained on the $D_{l} \cup D_{a}$. Until the ASR model $\theta_{asr}$ converges or reaches the desired performance, the active learning pipeline terminates.

\begin{algorithm}[ht] 
	\caption{Active Learning Algorithm} 
	\label{algorithm1} 
	\begin{algorithmic}[1]
		\STATE {\bfseries Input:} \\
		$D_{u}$: initially unlabeled dataset \\
		$D_{l}$: initially labeled dataset \\
		$K$: number of required training samples in each data selection\\
		\REPEAT
		\STATE train the ASR/LP joint model $[\theta_{asr},\theta_{lp}]$ on $D_{l}$ \\
		\STATE let selected dataset $D_{a}=\emptyset$\\
		\FOR {sample $x$ $\in$ $D_{u}$}
		\STATE compute rank metric $R_{x}=\theta_{lp}(x)$
		\ENDFOR
		\STATE sort sample $x$ in $D_{u}$ by $R_{x}$
		\FOR {$1$ to $K$}
		\STATE get $x^*=\mathop{\arg\max}(R_{x^*})$
		\STATE $D_{a} \leftarrow D_{a} \cup x^*$
		\STATE $D_{u} \leftarrow D_{u} \setminus x^*$
		\ENDFOR
		\STATE annotate $D_{a}$ by human experts
		\STATE $D_{l} \leftarrow D_{l} \cup D_{a}$
		\UNTIL {ASR model $\theta_{asr}$ reaches the desired performance}
		\RETURN $[\theta_{asr},\theta_{lp}]$
	\end{algorithmic} 
\end{algorithm}


%

\section{Experimental Setup}
\label{sec:exp}
In this section, we evaluate the proposed method in speech recognition tasks. We select the samples from the training set to simulate the active learning pipeline for annotation. We compare the performance of our approach with random selection and two other uncertainty-based AL methods. We also discuss the error function of predicted loss, and analyze the correlation about ranking metric.

\subsection{Dataset}
We performed the experiments on two open vocabulary datasets: (1) an English speech corpora, WSJ, (2) a Chinese speech corpus, AISHELL-1.
\begin{itemize}
	\setlength{\itemsep}{0pt}
	\setlength{\parsep}{0pt}
	\setlength{\parskip}{0pt}
	\item \textbf{WSJ}: The WSJ1~\cite{linguistic1994wsj1} and WSJ0~\cite{linguistic2007wsj0} are read speech on Wall Street Journal news text recorded by microphones. We use si284 for training, dev93 for validation, and eval92 for testing.
	\item \textbf{AISHELL-1}: The AISHELL-1~\cite{aishell_2017} was recorded in quiet indoor environment. The transcripts contain $11$ domains, including smart home, autonomous driving, and industrial production.
\end{itemize}

\begin{table}[ht]
	\centering
	\caption{Datasets of Speech Recognition Task}
	\label{tab1}
	\scalebox{0.9} {
	\begin{tabular}{p{3.5cm}|p{2cm}|p{2cm}}
		\hline\hline
		\textbf{WSJ} & \textbf{Utterances} & \textbf{Length (hours)} \\
		\hline
		Training 10\% & 3,741 & 8 \\
		Training 20\% & 7,482 & 16 \\
		Training 30\% & 11,223 & 24 \\
		Training 40\% & 14,963 & 32 \\
		Training 50\% & 18,704 & 40 \\
		Validation & 503 & 1.1 \\
		Testing & 333 & 0.7 \\
		\hline\hline
		\textbf{AISHELL-1} & \textbf{Utterances} & \textbf{Length (hours)} \\
		\hline
		Training 10\% & 12,010 & 15 \\
		Training 20\% & 24,020 & 30 \\
		Training 30\% & 36,030 & 45 \\
		Training 40\% & 48,040 & 60 \\
		Training 50\% & 60,049 & 75 \\
		Validation & 14,326 & 10 \\
		Testing & 7,176 & 5 \\
		\hline\hline
	\end{tabular}
    }
\end{table}

Table~\ref{tab1} summarizes the datasets of ASR task in the experiments. 
At each training iteration, top $K$ ranking samples $D_{a}$, equivalent to $10\%$ of the entire training set, are selected. $D_{a}$ were annotated and removed from $D_{u}$, then added to $D_{l}$ for the next model training.

\subsection{Model Configuration}

\begin{figure*}[ht]
	\centering
	\subfigure[WER on dev93(\%)] { \label{fig3a}
		\includegraphics[width=0.45\columnwidth]{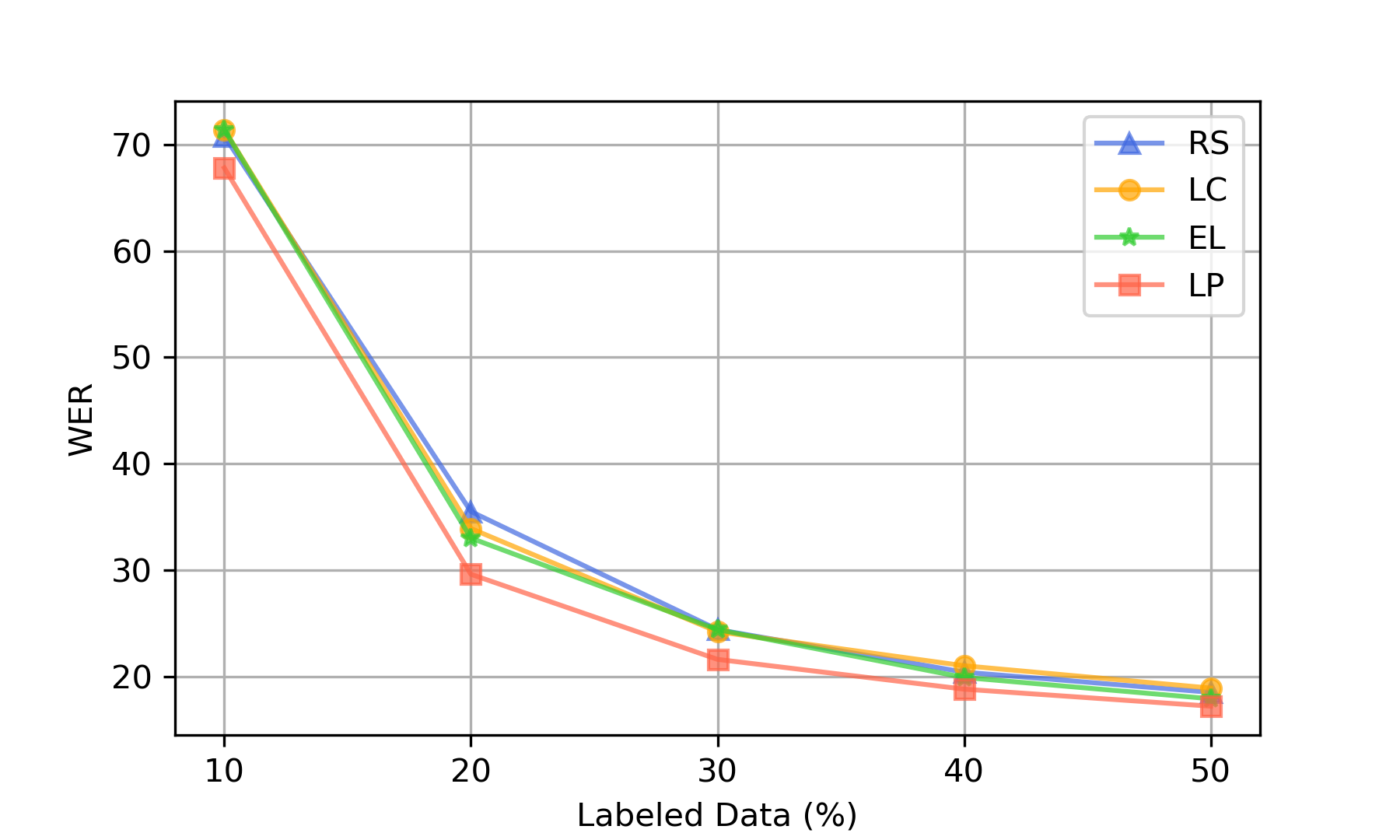}
	}
	\subfigure[Corr on dev93(\%)] { \label{fig3b}
		\includegraphics[width=0.45\columnwidth]{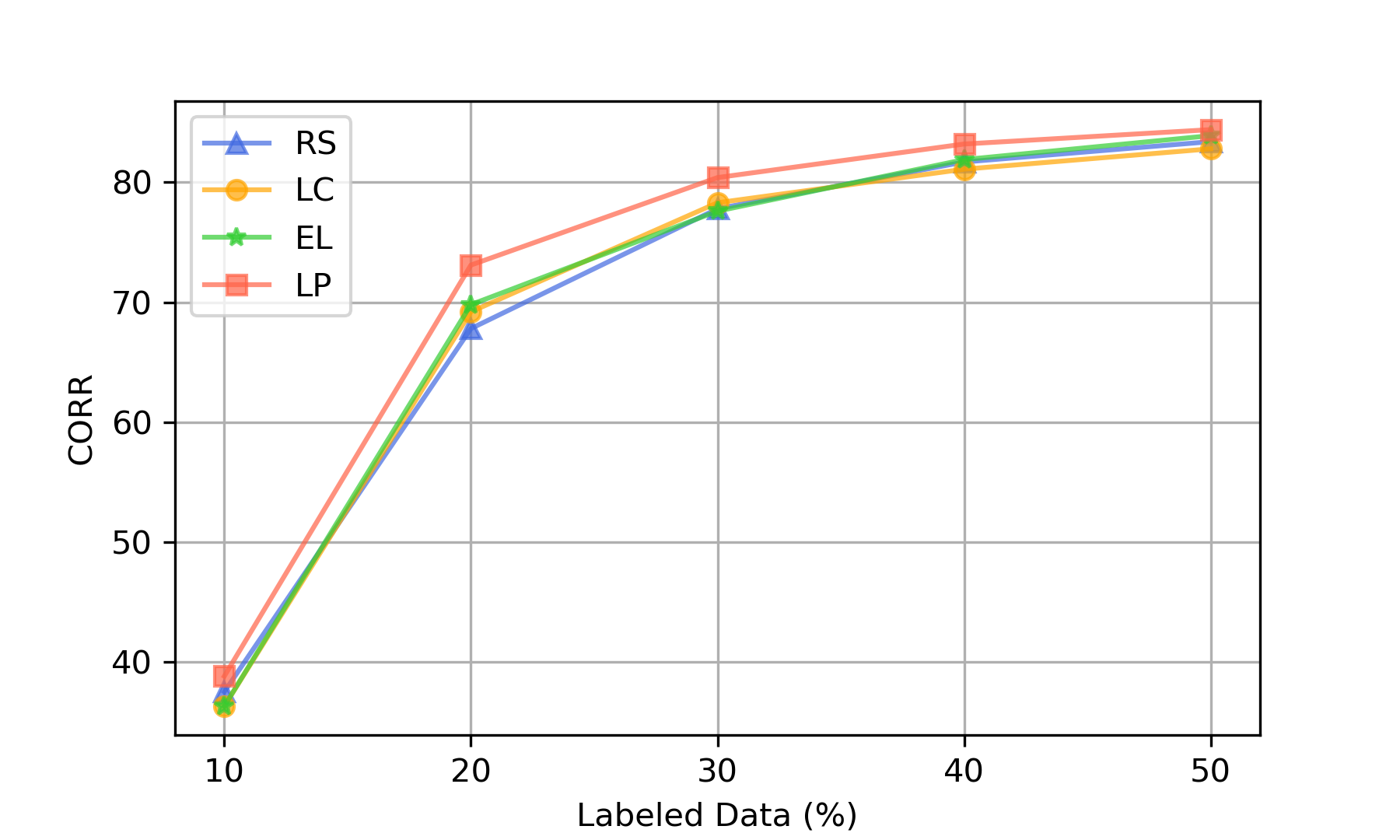}
	}
	\subfigure[WER on eval92(\%)] { \label{fig3c}
		\includegraphics[width=0.45\columnwidth]{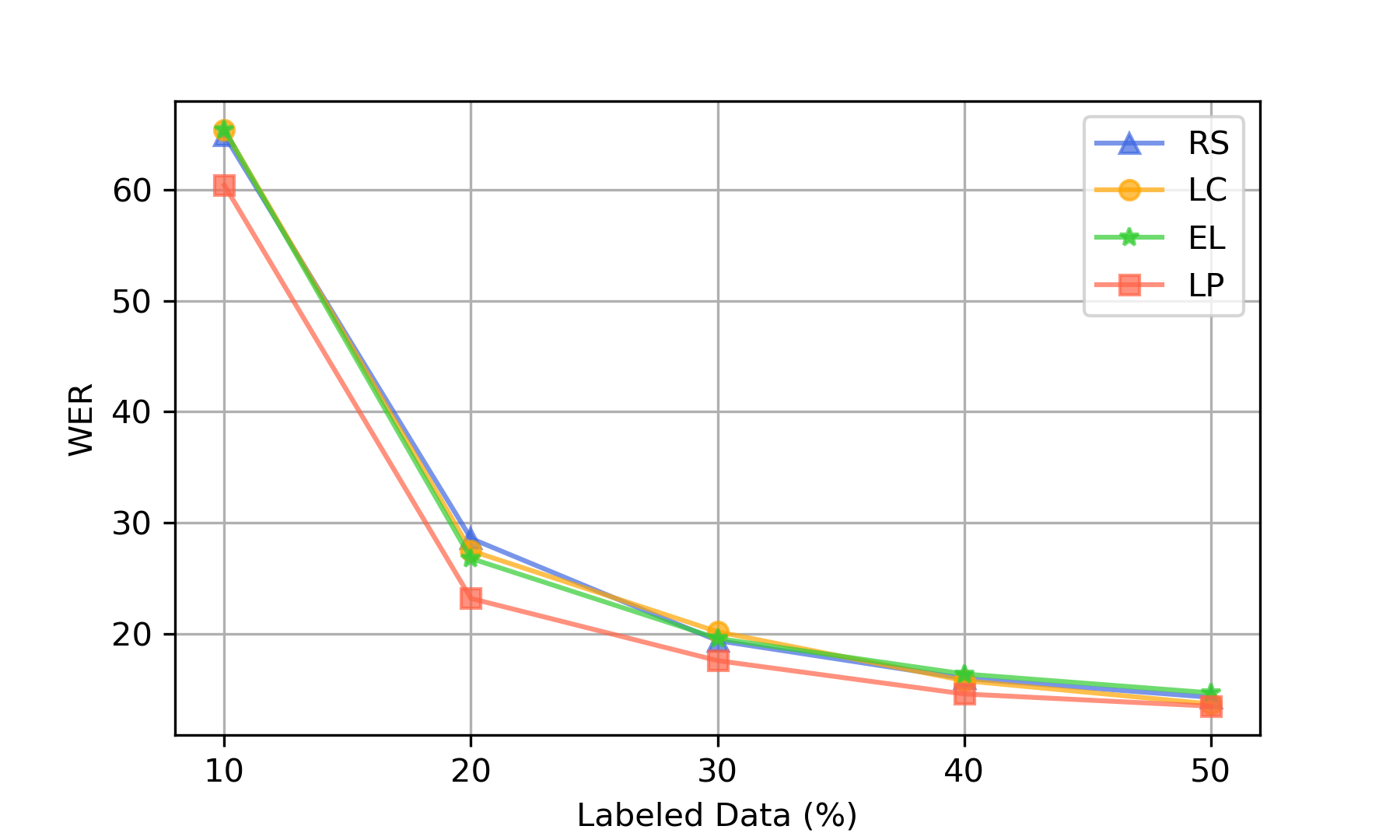}
	}
	\subfigure[Corr on eval92(\%)] { \label{fig3d}
		\includegraphics[width=0.45\columnwidth]{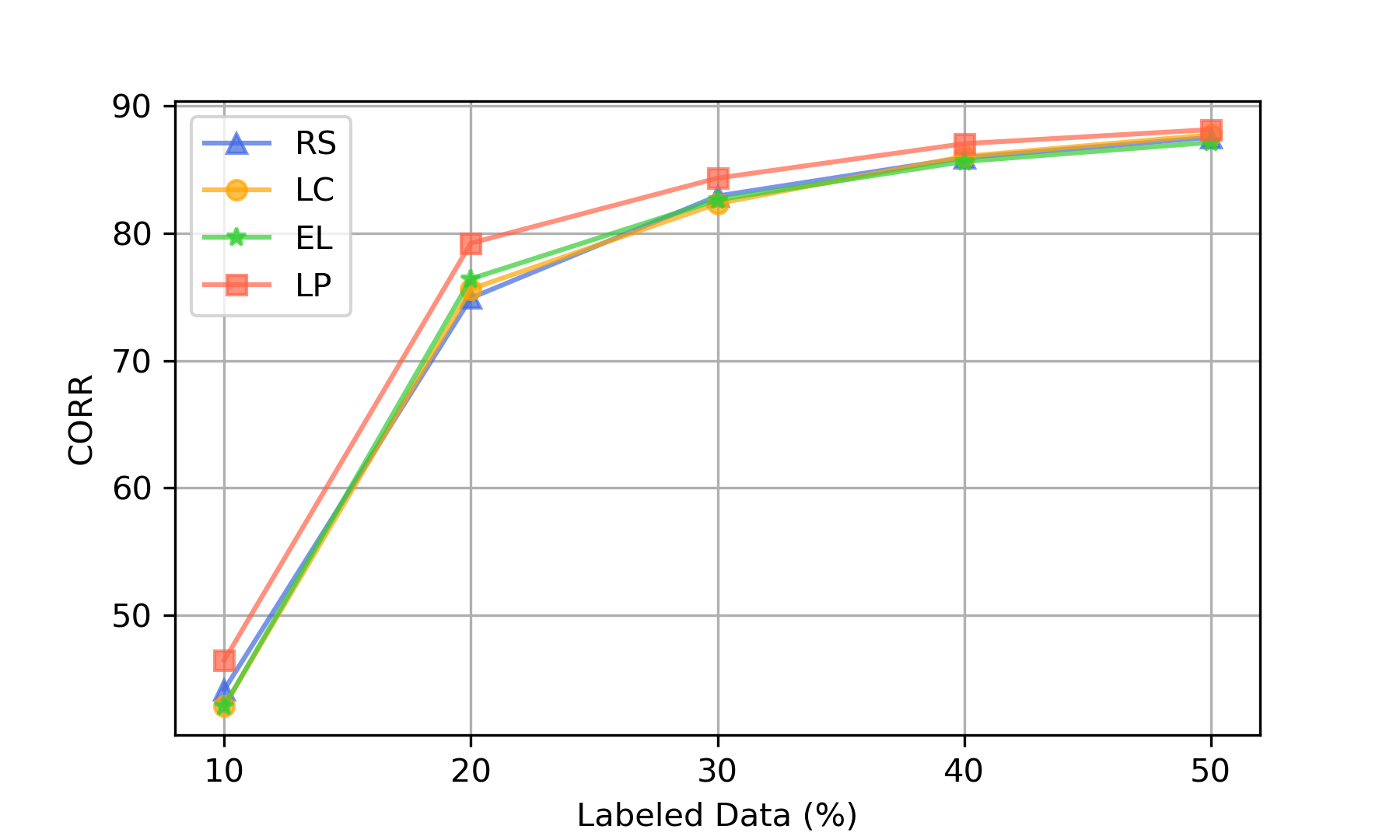}
	}
	\caption{Compared with Other Active Learning Methods, Results on WSJ}
	\label{fig3}
\end{figure*}

\begin{figure*}[ht]
	\centering
	\subfigure[CER on validation set(\%)] { \label{fig4a}
		\includegraphics[width=0.45\columnwidth]{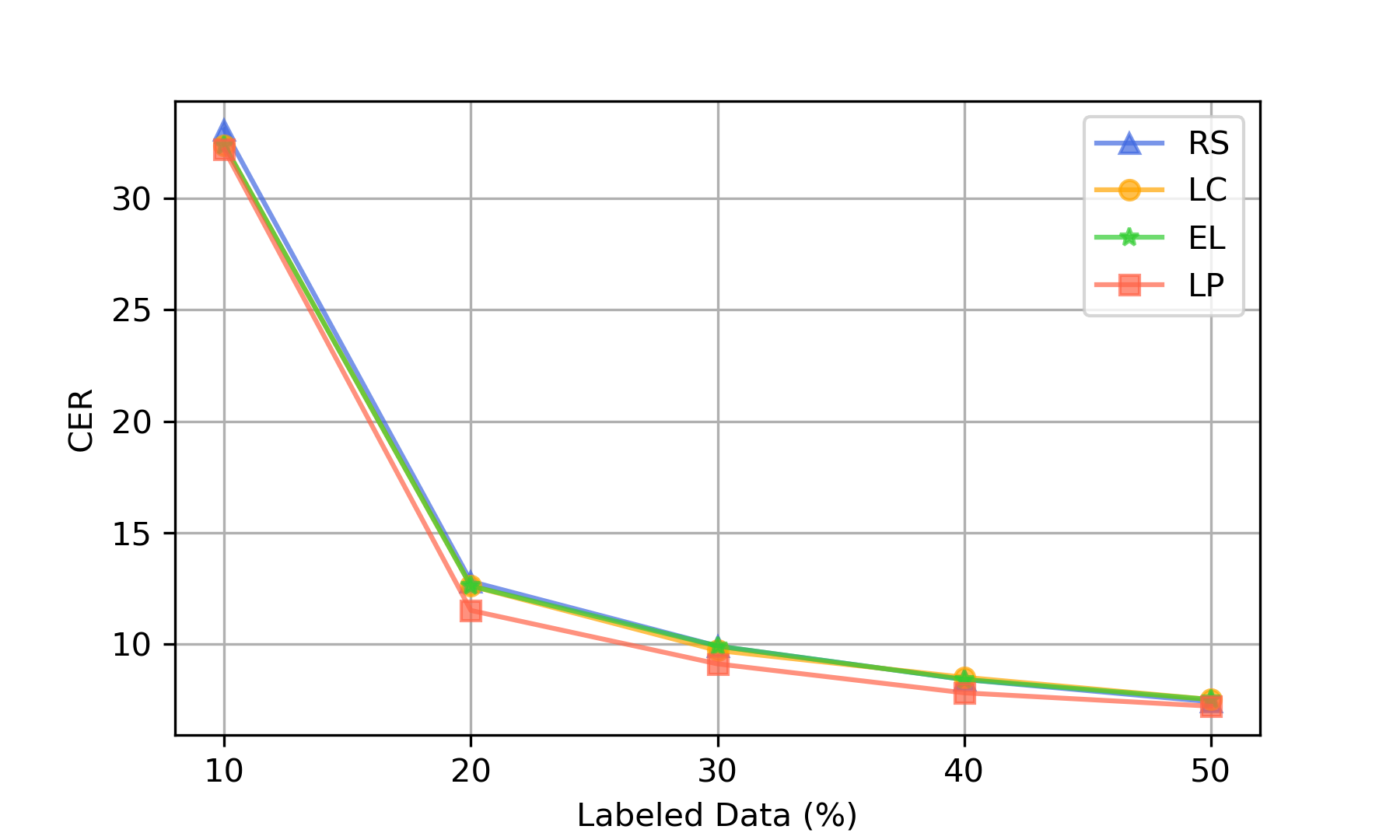}
	}
	\subfigure[Corr on validation set(\%)] { \label{fig4b}
		\includegraphics[width=0.45\columnwidth]{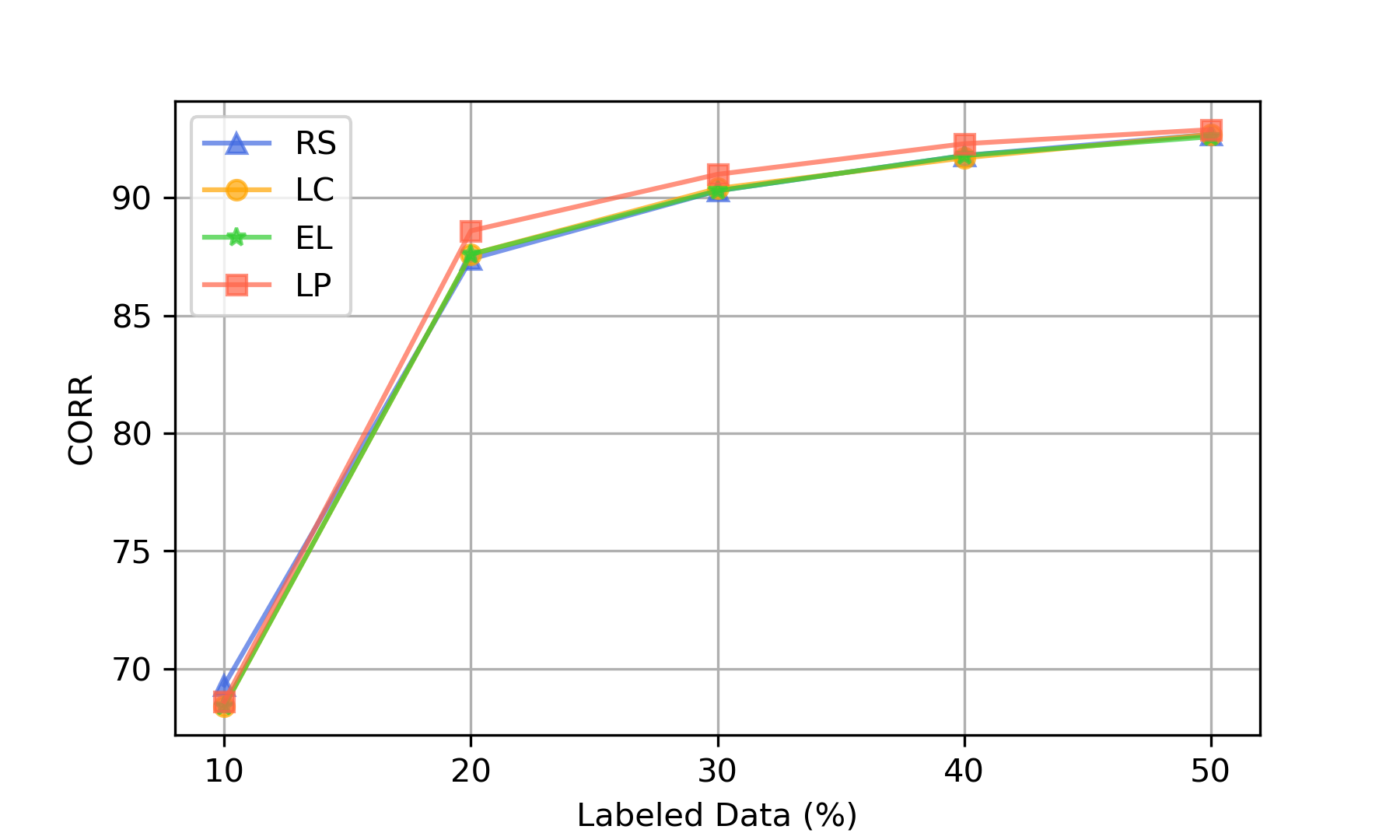}
	}
	\subfigure[CER on testing set(\%)] { \label{fig4c}
		\includegraphics[width=0.45\columnwidth]{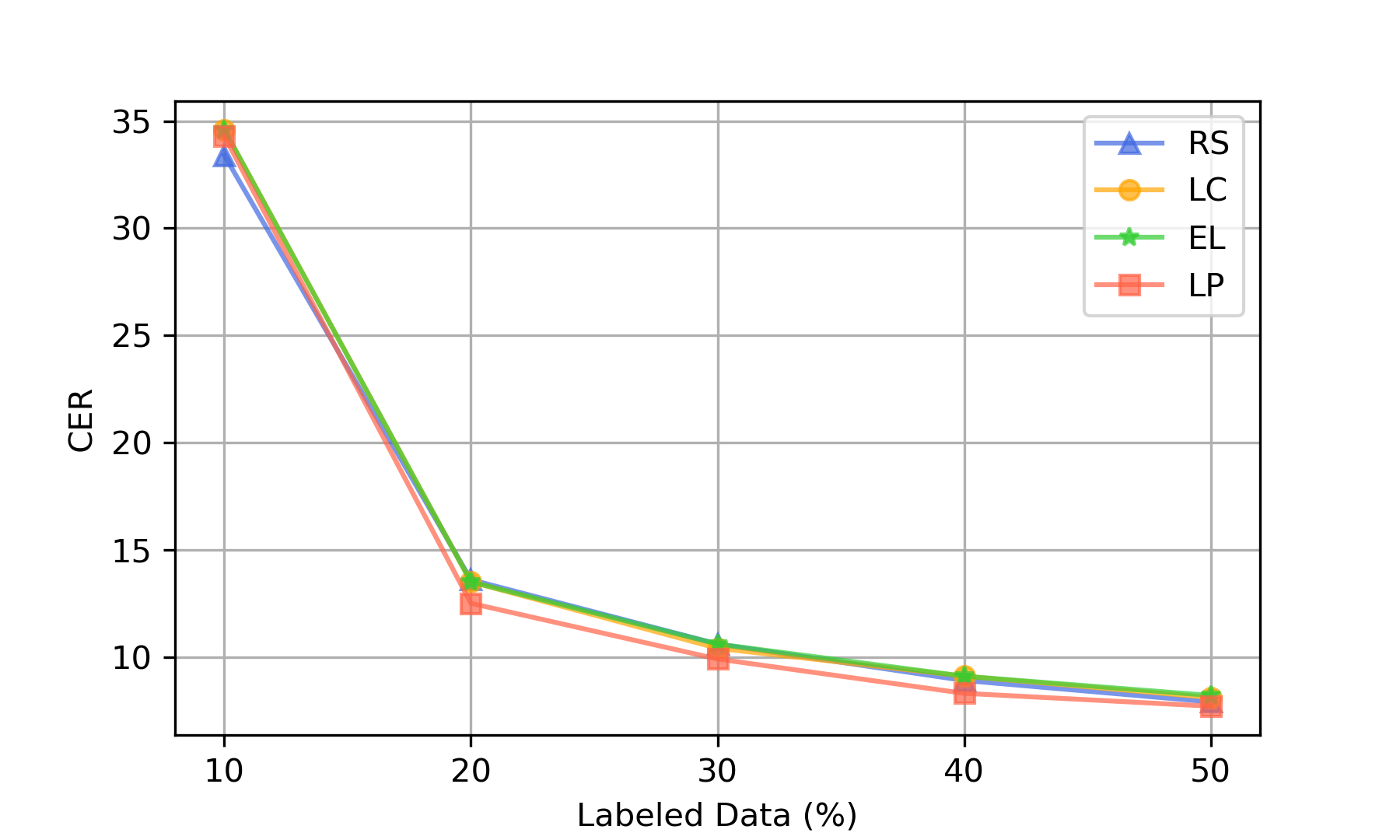}
	}
	\subfigure[Corr on testing set(\%)] { \label{fig4d}
		\includegraphics[width=0.45\columnwidth]{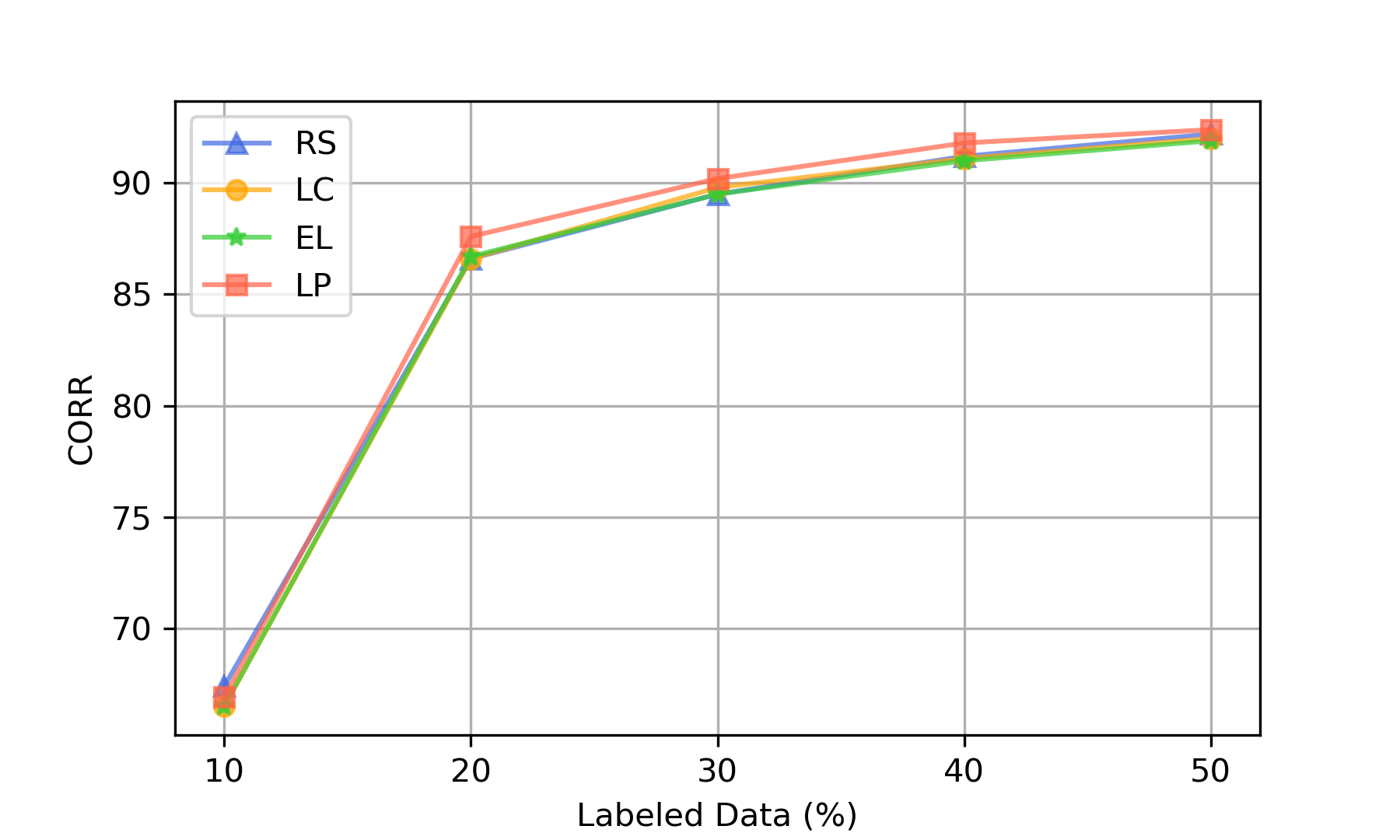}
	}
	\caption{Compared with Other Active Learning Methods, Results on AISHEEL-1}
	\label{fig4}
\end{figure*}

We implemented all the experiments using the ESPNET2 toolkit~\cite{watanabe2018espnet}, which based on Pytorch~\cite{paszke2019pytorch} deep learning framework. The parameters of ASR/LP joint model are listed in Table~\ref{tab2}. The audio data is encoded with $80$ Fbank coefficients of $25ms$ frame length and $10ms$ frame shift. The acoustic feature is firstly inputted into an encoder network $\theta_{encoder}$. $\theta_{encoder}$ consists of two-layer 2D convolutions, and twelve-layer transformer encoder blocks. The multi-head attention has $4$ heads, $256$ dimension, and the feedforward dimension is $2048$. Afterwards, the encoder output is fed into the CTC loss prediction network $\theta_{ctc}$. $\theta_{ctc}$ contains two-layer bidirectional LSTM layers, followed by two feedforward layers mapping from the features to the predicted CTC loss. Similarly, the decoder network $\theta_{decoder}$ is six-layer transformer decoder blocks, and outputs the probability of grapheme labels at each step. For WSJ, the dimension of decoder output is $65$ (including capital English letters, blank, and punctuations). For AISHELL-1, the decoder output is $4233$ dimension (including Chinese characters, symbols, and blank). The attention loss prediction network $\theta_{attention}$ takes the last layer hidden states of $\theta_{decoder}$ as input, followed by another two-layer bidirectional LSTM layers and feedforward layers to predict attention loss.

\begin{table}[ht]
	\centering
	\caption{Parameters of ASR/LP Joint Model}
	\label{tab2}
	\scalebox{0.9} {
	\begin{tabular}{p{4.3cm}|p{4.3cm}}
		\hline\hline
		\multicolumn{2}{c}{\textbf{Target Task Module $\theta_{asr}$}} \\
		\hline
		Encoder Network $\theta_{encoder}$ & Decoder Network $\theta_{decoder}$ \\
		\hline
		2$\times$Conv2d
		
		(dim=256, kernel=(3,3))
		&
		 \\
		 & \\
		12$\times$TransformerEncoder
		
		(dim=256, head=4, ffn=2048)
		&
		6$\times$TransformerDecoder
		
		(dim=256, head=4, ffn=2048) \\
		\hline\hline
		\multicolumn{2}{c}{\textbf{Loss Prediction Module $\theta_{lp}$}} \\
		\hline
		CTC Loss Predict $\theta_{ctc}$ & Attention Loss Predict $\theta_{attention}$ \\
		\hline
		2$\times$BLSTM
		
		(dim=256)
		&
		2$\times$BLSTM
		
		(dim=256) \\
		 & \\
		2$\times$FeedForward
		
		(dim=256, dim=1024)
		&
		2$\times$FeedForward
		
		(dim=256, dim=1024) \\
		\hline\hline
		\multicolumn{2}{l}{hyperparameter $\beta = 1, \lambda = 0.3$, $\mu = 0.1$}\\
		\hline\hline
	\end{tabular}
    }
\end{table}

Models in the experiments were trained on $4$ Nvidia V100 GPU cards by $100$ epochs for WSJ, and $2$ GPU cards by $100$ epochs for AISHELL-1.
Besides, the Adam optimizer~\cite{kingma2014adam} was used with an initial learning rate at $0.005$, and $30000$ warmup steps. The output model in each experiment is averaged from the best $10$ models according to the validation accuracy. 

\subsection{Results}
We compared our proposed \textbf{Loss Prediction (LP)} approach with random selection and two other uncertainty-based active learning methods:
\begin{itemize}
	\setlength{\itemsep}{0pt}
	\setlength{\parsep}{0pt}
	\setlength{\parskip}{0pt}
	\item \textbf{Random Selection (RS)}: randomly choosing samples for annotation. 
	\item \textbf{Least Confidence (LC)}: choosing samples with the least path probability of the top decoded path~\cite{malhotra2019active}.
	\item \textbf{Estimated Loss (EL)}: estimating the CTC and attention loss by most likely prediction as ground-truth, and choosing samples with top of these estimated loss~\cite{huang2016activelf}.
\end{itemize}

\begin{table}[ht]
	\centering
	\caption{Compared with Other AL Methods}
	\subtable[Results on eval92 of WSJ, WER/CORR (\%)]{
	\scalebox{0.9} {
	\begin{tabular}{p{1.0cm}|p{1.0cm}|p{1.2cm}|p{1.2cm}|p{1.2cm}|p{1.2cm}}
		\hline\hline
		\multicolumn{2}{c|}{\textbf{WSJ}} & \multicolumn{3}{c}{\textbf{WER on eval92}} \\
		\hline
		\textbf{Iteration} & \textbf{Training} & \textbf{RS} & \textbf{LC} & \textbf{EL} & \textbf{LP (ours)} \\
		\hline
		1 & 10\% & 64.9 & 65.4 & 65.4 & \textbf{60.4} \\
		2 & 20\% & 28.6 & 27.5 & 26.8 & \textbf{23.2} \\
		3 & 30\% & 19.4 & 20.2 & 19.6 & \textbf{17.6} \\
		4 & 40\% & 16.0 & 15.8 & 16.4 & \textbf{14.6} \\
		5 & 50\% & 14.3 & 13.7 & 14.7 & \textbf{13.5} \\
		\hline\hline
		\multicolumn{2}{c|}{\textbf{WSJ}} & \multicolumn{3}{c}{\textbf{CORR on eval92}} \\
		\hline
		\textbf{Iteration} & \textbf{Training} & \textbf{RS} & \textbf{LC} & \textbf{EL} & \textbf{LP (ours)} \\
		\hline
		1 & 10\% & 44.2 & 42.9 & 42.9 & \textbf{46.5} \\
		2 & 20\% & 74.9 & 75.6 & 76.4 & \textbf{79.2} \\
		3 & 30\% & 82.9 & 82.3 & 82.6 & \textbf{84.3} \\
		4 & 40\% & 85.9 & 86.0 & 85.6 & \textbf{87.0} \\
		5 & 50\% & 87.5 & 87.7 & 87.1 & \textbf{88.1} \\
		\hline\hline
	\end{tabular}
    }
    }
    \qquad
    \subtable[Results on testing set of AISHELL-1, CER/CORR (\%)]{
    \scalebox{0.9} {
	\begin{tabular}{p{1.0cm}|p{1.0cm}|p{1.2cm}|p{1.2cm}|p{1.2cm}|p{1.2cm}}
		\hline\hline
		\multicolumn{2}{c|}{\textbf{AISHELL-1}} & \multicolumn{3}{c}{\textbf{CER on testing set}} \\
		\hline
		\textbf{Iteration} & \textbf{Training} & \textbf{RS} & \textbf{LC} & \textbf{EL} & \textbf{LP (ours)} \\
		\hline
		1 & 10\% & \textbf{33.4} & 34.6 & 34.6 & 34.3 \\
		2 & 20\% & 13.6 & 13.5 & 13.5 & \textbf{12.5} \\
		3 & 30\% & 10.6 & 10.4 & 10.6 & \textbf{9.9} \\
		4 & 40\% & 8.9 & 9.1 & 9.1 & \textbf{8.3} \\
		5 & 50\% & 7.9 & 8.1 & 8.2 & \textbf{7.7} \\
		\hline\hline
		\multicolumn{2}{c|}{\textbf{AISHELL-1}} & \multicolumn{3}{c}{\textbf{CORR on testing set}} \\
		\hline
		\textbf{Iteration} & \textbf{Training} & \textbf{RS} & \textbf{LC} & \textbf{EL} & \textbf{LP (ours)} \\
		\hline
		1 & 10\% & \textbf{67.4} & 66.5 & 66.5 & 66.9 \\
		2 & 20\% & 86.6 & 86.6 & 86.7 & \textbf{87.6} \\
		3 & 30\% & 89.5 & 89.8 & 89.5 & \textbf{90.2} \\
		4 & 40\% & 91.2 & 91.1 & 91.0 & \textbf{91.8} \\
		5 & 50\% & 92.2 & 92.0 & 91.9 & \textbf{92.4} \\
		\hline\hline
	\end{tabular}
    }
    }
\label{tab3}
\end{table}

\begin{figure*}[ht]
	\centering
	\subfigure[WER(\%)] { \label{fig5a}
		\includegraphics[width=0.55\columnwidth]{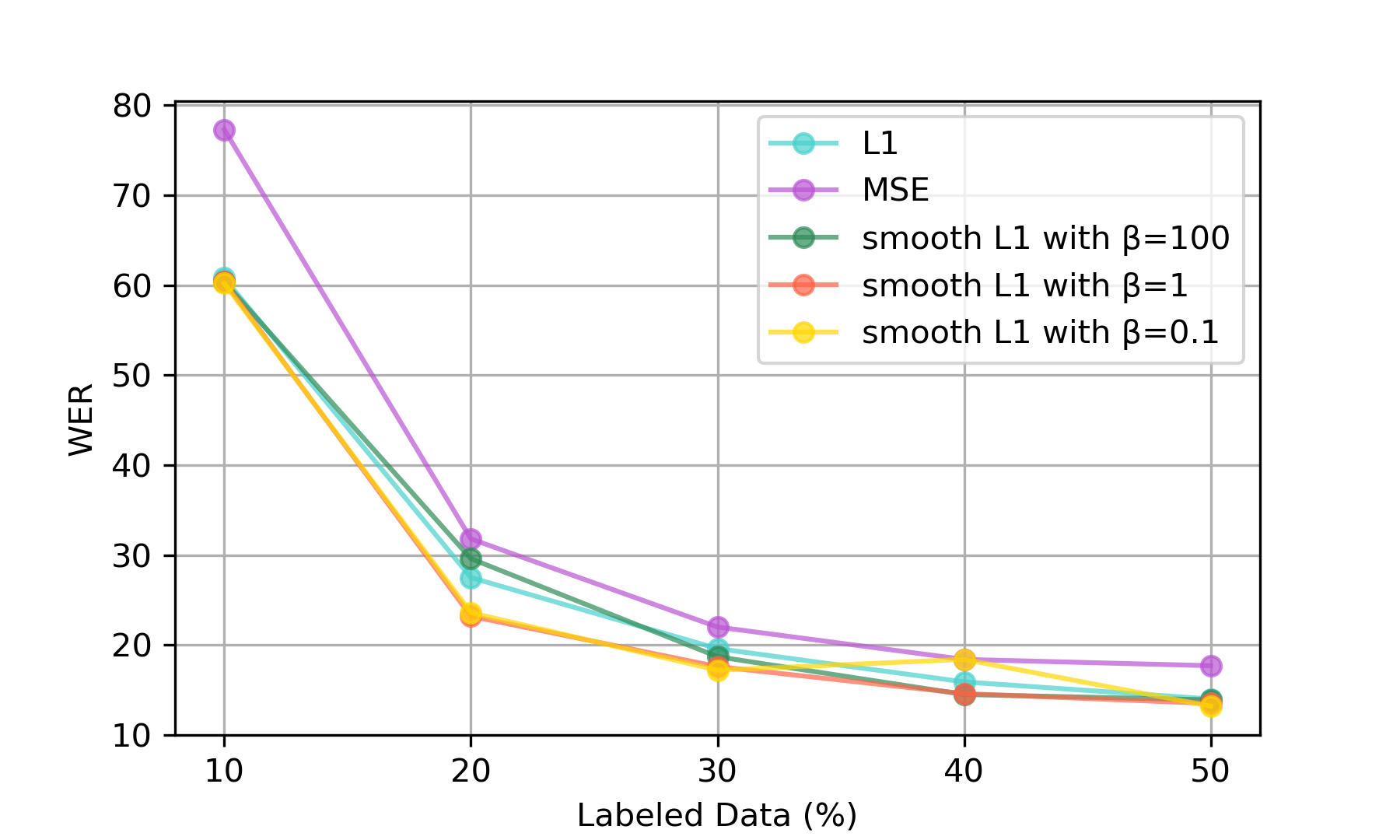}
	}
	\subfigure[Corr(\%)] { \label{fig5b}
		\includegraphics[width=0.55\columnwidth]{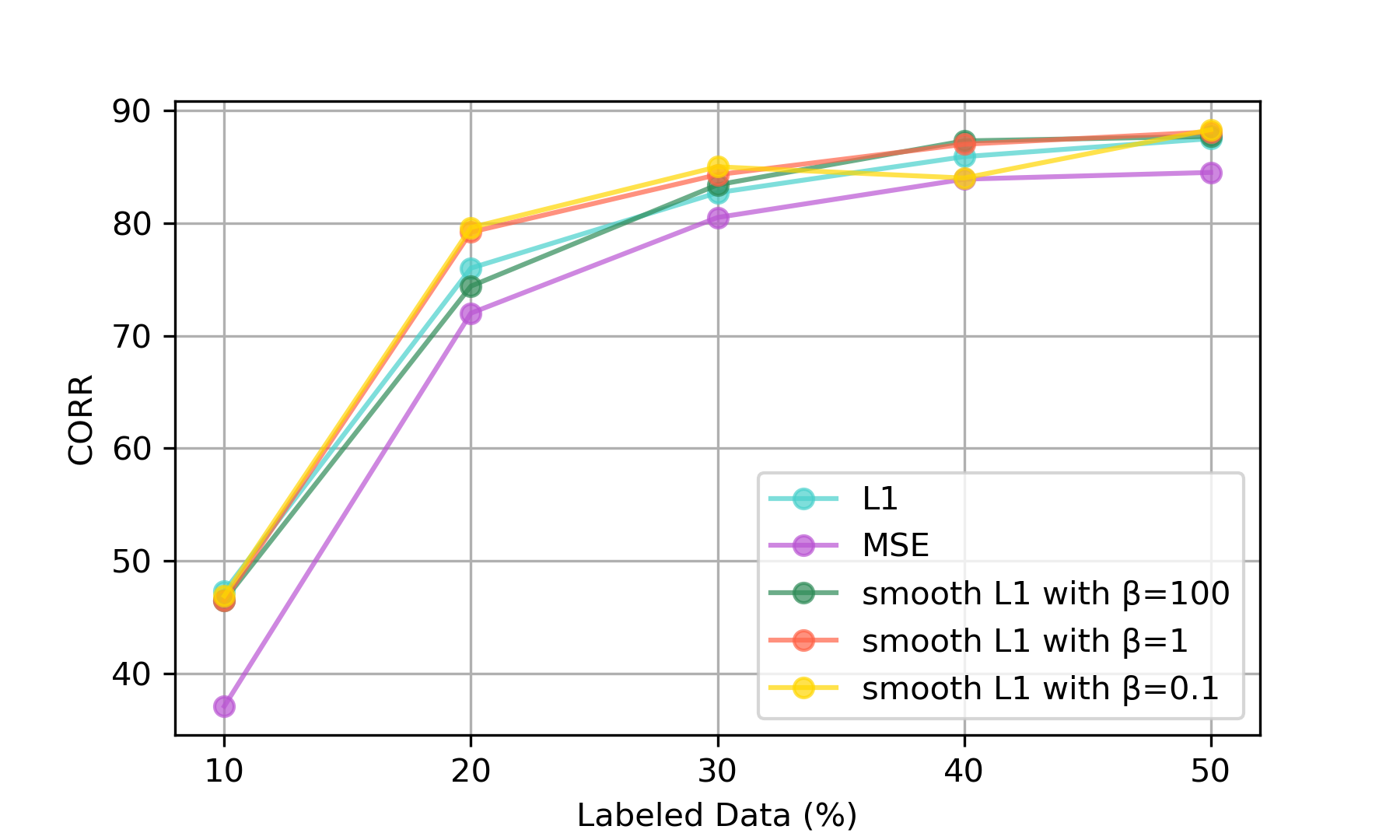}
	}
	\caption{Ablation Study, Different Choices on Error Functions, eval92 of WSJ}
	\label{fig5}
\end{figure*}

\begin{figure*}[ht]
	\centering
	\subfigure[Ranking Metric vs. CER] { \label{fig6a}
		\includegraphics[width=0.55\columnwidth]{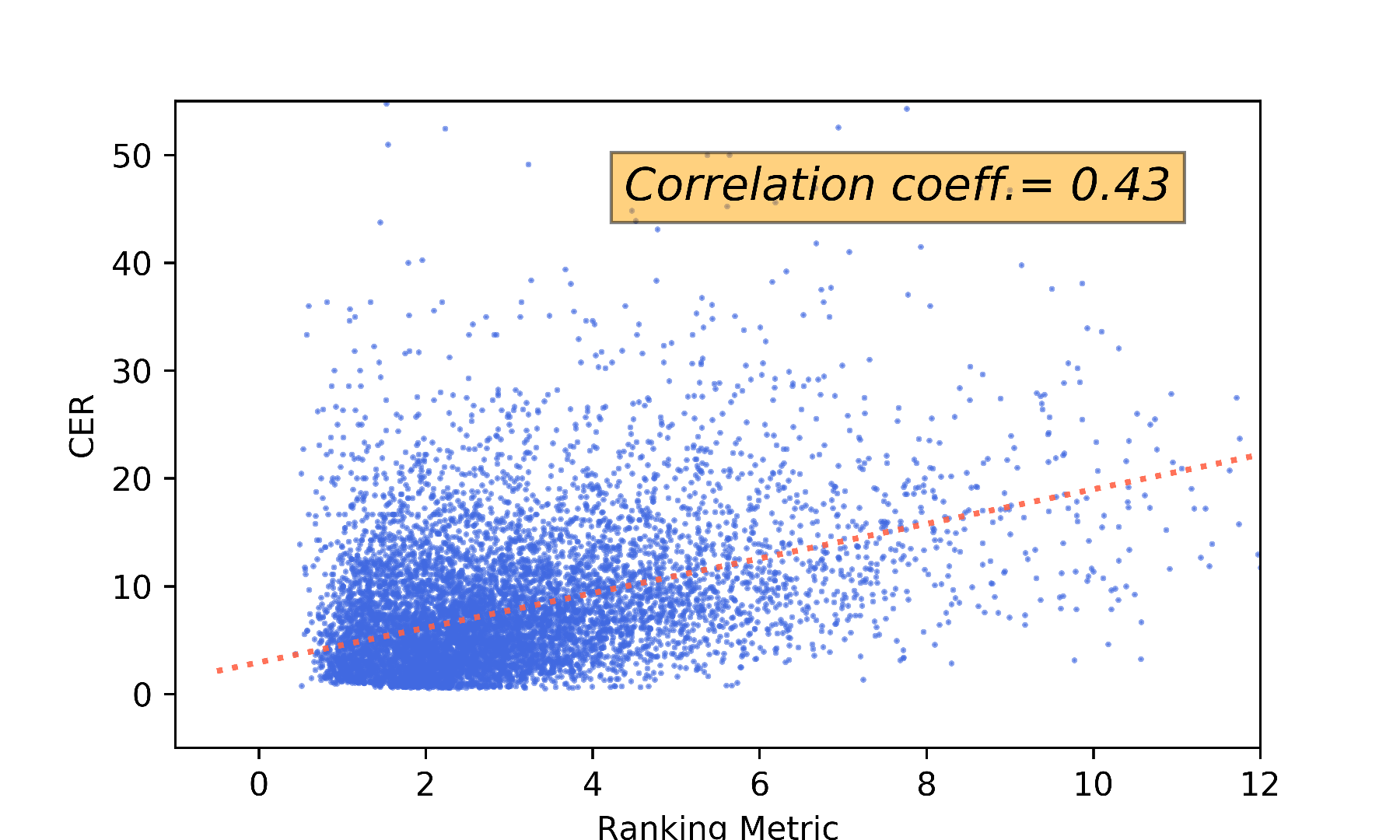}
	}
	\subfigure[True Loss vs. Ranking Metric] { \label{fig6b}
		\includegraphics[width=0.55\columnwidth]{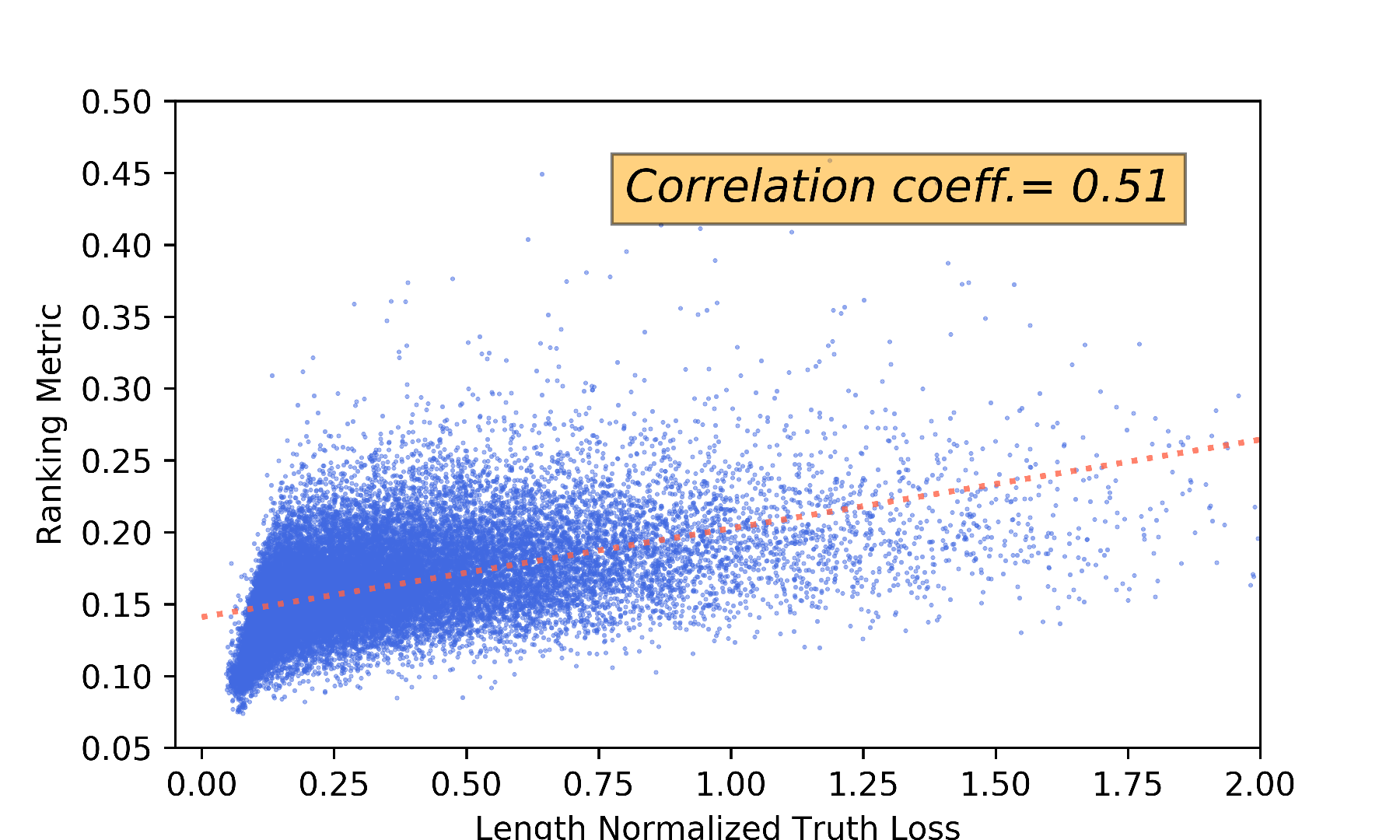}
	}
	\caption{The Analysis of Ranking Metric}
	\label{fig6}
\end{figure*}

The LP approach was training a ASR/LP joint model, while the RS and LC method only used the ASR model. Fig~\ref{fig3} illustrates Word Error Rate (WER) and Word Correct Rate(CORR) results on WSJ, and Fig~\ref{fig4} illustrates Character Error Rate (CER) and CORR results on AISHELL-1.
To investigate the performance difference, all the models were evaluated without applying the language model. The results infer that our proposed LP approach has the best performance, compared with other uncertainty-based LC and EL methods. In addition, both the LP and LC approach outperform the RS method. Table~\ref{tab3} lists all the results for $5$ training iterations on WSJ and AISHELL-1. On the $20\%$ labeled data, the LP approach achieves $18.9\%$ relative improvement on WER of WSJ, and $8.1\%$ on CER of AISHELL-1, over the RS method. During the first iteration of AISHELL-1 training, we found that the RS approach works best. We conjecture that it might need more diversity of the samples in the early training stage of chinese ASR model. As the training data increases, the LP approach outperforms other methods at resting experiments.

\subsection{Analysis}
\label{sec:analysis}
We investigated different choices on the error function of predicted loss. Five error functions were analyzed: (1) L1 Loss, (2) MSE Loss, (3) SmoothL1Loss with $\beta=100$, (4) SmoothL1Loss with $\beta=1$, and (5) SmoothL1Loss with $\beta=0.1$. CTC error $\mathcal{E}_{ctc}$ and attention error $\mathcal{E}_{attention}$ were used the same error function. As Fig~\ref{fig5} depicts, the SmoothL1Loss provides significant improvements over L1 and MSE Loss on WSJ. The hyperparameter $\beta$ controls the smooth degree of the error function. When $\beta$ is set to $0$, it degrades to L1 Loss. In our experiments, we find that the ASR/LP joint model has the best performance when $\beta=1$.

In Fig~\ref{fig6a}, we visualized the relationship between the CER and the ranking metric $R_x$, which is calculated by the unlabeled dataset $D_u$ of the $20\%$ active learning cycle on WSJ. The correlation coefficient is $0.43$, meaning that the data samples selected by our LP algorithm also have high CER values. Fig~\ref{fig6b} depicts the scatter plot between the ranking metric $R_x$ and the length-normalized true loss $\mathcal{L}$, computed on $20\%$ training iteration of AISHELL-1. For comparison, the true loss $\mathcal{L}$ is also normalized by the length $T$ and $S$, similar to Eq.~\ref{rank_metric}. Two values are correlated with the correlation coefficient of $0.51$. The result indicates that the LP module fits the loss well. Overall, the visualization demonstrates that the LP approach is an effective active learning method of selecting informative data samples. Under the same training cost, the ASR model with the LP approach can obtain a substantial performance improvement.

\section{Conclusions}
\label{sec:conclu}
In this paper, we proposed an end-to-end active learning approach for speech recognition tasks. The active learning approach uses an ASR/LP joint model. The ASR module conducts the speech recognition task, and the LP module predicts the CTC and attention loss. We choosed SmoothL1Loss as the error function of LP module training. The ranking metric for selecting samples is computed by length-normalized predicted loss. The experiments demonstrate that our approach outperforms the random selection and other uncertainty-based methods on WSJ and AISHELL-1 datasets. Future works, including taking data diversity into consideration and designing a better ranking metric, can be extended. We are also interested in combining the active learning method with the semi-supervised learning on unlabeled data, providing further improvements on the training efficiency of ASR models.

\section{Acknowledgement}
\label{sec:ack}
This paper is supported by National Key Research and Development Program of China under grant No. 2018YFB0204403 , No. 2017YFB1401202 and No. 2018YFB1003500. Corresponding author is Jianzong Wang from Ping An Technology (Shenzhen) Co., Ltd.

\vfill\pagebreak
\bibliographystyle{IEEEtran}
\bibliography{IJCNN2021_Active_Learning}

\end{document}